\documentclass[
reprint,
superscriptaddress,
preprintnumbers,
 amsmath,
 amssymb,
 prc,
]{revtex4-1}

\usepackage[     bookmarks,
                 bookmarksopen = true,
                 bookmarksnumbered = true,
                 linktocpage,
                 colorlinks = true,
                 linkcolor = blue,
                 urlcolor  = blue,
                 citecolor = blue,
                 anchorcolor = green,
                 hyperindex = true,
                 hyperfigures]
        {hyperref}

\usepackage{graphicx}
\usepackage{dcolumn}
\usepackage{bm}
\usepackage{float}
\usepackage{amsmath}
\usepackage{soul}
\usepackage{color}
\usepackage{graphicx}
\definecolor{dgreen}{cmyk}{1.,0.,1.,0.2}        
\definecolor{orange}{cmyk}{0.,0.353,1.,0.}    

\begin{document}



\title{Four-dimensional QCD equation of state from a quasi-parton model with physics-informed neural networks}

\author{Fu-Peng Li}
\affiliation{Key Laboratory of Quark and Lepton Physics (MOE) \& Institute of Particle Physics, Central China Normal University, Wuhan 430079, China}
\affiliation{Artificial Intelligence and Computational Physics Research Center, Central China Normal University, Wuhan 430079, China}
\affiliation{Key Laboratory of Nuclear Physics and Ion-beam Application (MOE) \& Institute of Modern Physics, Fudan University, Shanghai 200433, China}
\affiliation{Shanghai Research Center for Theoretical Nuclear Physics,
NSFC and Fudan University, Shanghai 200438, China}

\author{Long-Gang Pang}
\email[]{lgpang@ccnu.edu.cn}
\affiliation{Key Laboratory of Quark and Lepton Physics (MOE) \& Institute of Particle Physics, Central China Normal University, Wuhan 430079, China}
\affiliation{Artificial Intelligence and Computational Physics Research Center, Central China Normal University, Wuhan 430079, China}

\author{Guang-You Qin}
\email[]{guangyou.qin@ccnu.edu.cn}
\affiliation{Key Laboratory of Quark and Lepton Physics (MOE) \& Institute of Particle Physics, Central China Normal University, Wuhan 430079, China}
\affiliation{Artificial Intelligence and Computational Physics Research Center, Central China Normal University, Wuhan 430079, China}

\date{\today}%

\begin{abstract}

The equation of state (EoS) of strongly interacting matter at finite temperature and chemical potentials (baryon, charge, and strangeness) is a crucial input for hydrodynamic simulations of relativistic heavy-ion collisions. We construct a four-dimensional EoS using a deep-learning-assisted quasi-particle model (DLQPM) within a physics-informed neural network (PINN) framework, in which the masses of light quarks, strange quarks, and gluons are parameterized as functions of temperature and chemical potentials ($T, \mu_B, \mu_Q, \mu_S$). The model is constrained by lattice QCD data at vanishing chemical potentials and provides a thermodynamically consistent extrapolation to finite $\mu_{B,Q,S}$. The DLQPM accurately reproduces the lattice-calculated cumulants $\chi^{B,Q,S}_{i,j,k}$ at $\mu_{B,Q,S}=0$, and its predicted EoS at various chemical potentials agrees well with results from the generalized $T'$-expansion method in lattice QCD. Furthermore, the calculated baryon-strangeness correlation $C_{BS}$ is consistent, within uncertainties, with preliminary STAR data. This work offers a reliable EoS for exploring the QCD phase structure in the beam energy scan region.
\end{abstract}
\maketitle

\section{Introduction}
A four-dimensional quantum chromodynamics equation of state (4D QCD EoS) is essential for investigating the properties of QCD matter under extreme conditions\cite{Shen:2014vra,Pang:2012he,Pang:2018zzo,Bernhard:2019bmu,Achenbach:2023pba,Pang:2016vdc,Braun-Munzinger:2015hba,Elfner:2022iae,Monnai:2019hkn,Monnai:2021kgu,Denicol:2018wdp,Sorensen:2023zkk,An:2021wof,MUSES:2023hyz,Wu:2021fjf}. The EoS not only enhances our understanding of QCD thermodynamics across varying chemical potentials but also provides crucial input for relativistic hydrodynamic simulations. Recent progress in lattice QCD (LQCD) has yielded precise determinations of the QCD EoS at zero baryon chemical potential ($\mu_B = 0$), along with extensive susceptibility data ($\chi_{i,j,k}^{B,Q,S}$) under the same conditions\cite{Borsanyi:2013bia,HotQCD:2018pds,HotQCD:2012fhj,HotQCD:2014kol,HotQCD:2018pds,Bollweg:2022rps,Bollweg:2021vqf}. Building on these LQCD results, researchers have employed various expansion techniques---such as Taylor expansion and the $T^\prime$-expansion---to extend the EoS into the four-dimensional parameter space, where the $T^\prime$-expansion reorganizes the chemical-potential dependence into an effective temperature shift along lines of approximately constant density, thereby improving convergence and extending the reliable reach to larger chemical potential \cite{Mondal:2021jxk,Borsanyi:2021sxv,Borsanyi:2022qlh,Monnai:2024pvy,Abuali:2025tbd}. In contrast to these conventional approaches, our work introduces a 4D deep-learning quasi-parton model (DLQPM) to explore the complete 4D QCD EoS.

Effective models based on quasiparticle theory offer a practical and physically transparent framework for describing QCD matter by encoding strong-interaction effects into temperature- and chemical-potential-dependent quasiparticle properties. Among these, quasiparticle models employing the hard-thermal-loop (HTL) parametrization have been widely used to compute transport properties of the quark-gluon plasma (QGP), such as the squared speed of sound ($c_s^2$), bulk viscosity ($\zeta$), shear viscosity ($\eta$), and heavy-quark diffusion coefficients ($D_s$) \cite{Soloveva:2023tvj,Kang:2022jbg,Peshier:1995ty,Alqahtani:2015qja,Liu:2021dpm,Soloveva:2021quj,Plumari:2011mk,Li:2022ozl,Li:2025csc,Mykhaylova:2020pfk,Romatschke:2011qp,2653187}. However, relying on specific parametrization ans\"{a}tze inevitably introduces prior biases and simultaneously restricts the model's representational flexibility, making such approaches insufficient for describing quasiparticle systems in a four-dimensional thermodynamic parameter space. To address these limitations, we propose a 4D DLQPM. In the DLQPM, deep neural networks (DNNs) are used to represent effective quasiparticle masses, while automatic differentiation techniques are applied to consistently derive the EoS within a statistical thermodynamic framework \cite{thuerey2021pbdl,NEURIPS2019_bdbca288,JMLR:v18:17-468,Hornik1989MultilayerFN,1976A,Kapusta:2006pm}. This machine-learning approach, in which neural networks represent unknown functions and physical laws are incorporated directly as constraints during training, is known as physics-informed neural networks (PINNs) \cite{RAISSI2019686,2018PINN}. PINNs have been widely applied to solving numerous inverse problems in nuclear physics \cite{Wang:2024dzc,Soma:2023fyi,Pang:2024kid,Shi:2021qri,Zhou:2023pti,Aarts2025,Soma:2023qax}.

The RHIC-STAR experiment has recently released preliminary results on measurements of the baryon-strangeness correlation coefficient $C_{BS}$ at various beam energies. The $C_{BS}$ characterizes the relationship between fluctuations in baryon number and strangeness, thereby providing a crucial diagnostic observable for discriminating among various phases of strongly interacting matter \cite{Koch:2005vg,Koch:2006ek,STAR-QM,Luo:2017faz,Cheng:2008zh,Bazavov:2014xya,Bazavov:2013dta,Bollweg:2024epj,Koch:2008ia,Koch:2025cog}. 
Theoretically, the $C_{BS}$ can be calculated in the following\cite{Koch:2005vg,Koch:2006ek}:
\begin{align}
C_{BS} = -3 \frac{\langle BS\rangle}{\langle S^2\rangle} = -3\chi^{BS}_{11}/\chi^{SS}_{2}.
\label{eq:CBS}
\end{align}
In experiment or simulation, the $C_{BS}$ can be measured by the final state hadrons following:
\begin{align}
C_{BS} \approx  3\frac{\langle \Lambda\rangle+\langle \overline \Lambda\rangle + ...+ 3\langle\Omega^-\rangle + 3\langle\overline \Omega^+\rangle}{\langle K^0\rangle + \langle\overline K^0\rangle + ... + 9 \langle\Omega^-\rangle + 9\langle\overline\Omega^+\rangle}.
\end{align}
This framework enables both LQCD and transport models to perform quantitative calculations of the baryon–strangeness correlation $C_{BS}$. LQCD describes the preliminary STAR measurements for central collisions reasonably well at beam energies above 39 GeV. At lower energies, however, noticeable discrepancies between LQCD predictions and experimental data emerge \cite{Bollweg:2024epj}, in contrast to the trends obtained with the hadronic transport model UrQMD \cite{STAR-QM}. These deviations at low beam energies may arise from several sources, including the limited convergence of Taylor expansions at large $\mu_B$ in LQCD, ambiguities in relating experimental observables to theoretical conserved-charge cumulants, and possible non-equilibrium effects, medium modifications, or critical fluctuations that become increasingly relevant in low-energy collisions. Given that the experimental measurements are still preliminary, these comparisons should be interpreted with due caution.

The remainder of this paper is organized as follows. In Sec.~\ref{Met}, we present the DLQPM, including the DNN architecture, input data, and the construction of the loss function. The numerical results, including the equation of state and the baryon–strangeness correlation $C_{BS}$, are discussed in Sec.~\ref{Res}. Finally, conclusions and perspectives are given in Sec.~\ref{Sum}.

\begin{figure*}[htb]
\begin{centering}
\includegraphics[width=0.8\textwidth]{PINN_new.pdf}
\caption{(Color online) The input layer comprises four dimensions, specifically $T$, $\mu_B$, $\mu_Q$ and $\mu_S$. The three mass models present $u/d$, $s$ and gluon, respectively. Each mass model is characterized by a neural network with identical architecture, where the corresponding network parameters are denoted as $\theta_1$, $\theta_2$ and $\theta_3$. The output will undergo numerical integration to compute the partition function, $\ln Z$, followed by automatic differentiation to derive the corresponding thermodynamic quantities.}
\label{fig:network}
\end{centering}
\end{figure*}

\section{Method} 
\label{Met}
PINNs employ DNNs to approximate unknown functions while incorporating physical constraints during training. This approach has been widely applied in nuclear physics. The DLQPM is a specific implementation of PINNs.
As illustrated in Fig.~\ref{fig:network}, the model describes quasi-parton masses as functions of temperature ($T$), baryon chemical potential ($\mu_B$), charge chemical potential ($\mu_Q$), and strangeness chemical potential ($\mu_S$). Three distinct mass models are used to represent the quasi-parton masses. Each ResNet consists of 8 hidden layers, with 32 neurons per layer~\cite{ramachandran2017searchingactivationfunctions}.
These mass model outputs are then used to compute the total partition function $\ln Z(T, \mu_B, \mu_Q, \mu_S)$~\cite{Kapusta:2006pm}:
\begin{align}
\ln Z(T,\mu_B,\mu_Q,\mu_S) & = \ln Z_g(T,\mu_B,\mu_Q,\mu_S) \nonumber \\
& + \sum_i \ln Z_{q(\bar{q})_i}(T,\mu_B,\mu_Q,\mu_S),
\label{eq:lnz}
\end{align}
where $i$ denotes the quark flavor. We adopt the assumption that quark and antiquark masses are identical. The explicit form of the partition function  $\ln Z(T,\mu_B,\mu_Q,\mu_S)$ can be expressed through the ideal gas statistical formulation:
\begin{widetext}
\begin{align}
\ln Z_g(T,\mu_B,\mu_Q,\mu_S) &= - \frac{d_g V}{2 \pi^{2}} \int_{0}^{\infty}p^{2} dp  \ln \left[ 1 - \exp \left(-{1 \over T}\sqrt{p^{2}+m_g^{2}}\right) \right], \nonumber \\
\ln Z_{q(\bar{q})_i}(T,\mu_B,\mu_Q,\mu_S) &= + \frac{d_{q(\bar{q})_i} V}{2 \pi^{2}} \int_{0}^{\infty}p^{2} dp \ln \left[ 1 + \exp \left(- {1 \over T}\left(\sqrt{p^{2}+m_{q(\bar{q})_i}^{2}} \pm \mu_{q(\bar{q})_i}\right)\right) \right],
\label{eq:lnz_gluon}
\end{align}
\end{widetext}
$d_g$ and $d_{q(\bar{q})_i}$ denote the degeneracies of the gluon and the quarks (antiquarks). In this calculation, $d_g = 2 \times 8 = 16$ and $d_{q(\bar{q})_i} = 2 \times 3 = 6$. The quark chemical potentials are given by\cite{HotQCD:2012fhj}:
\begin{align}
\mu_u &= \frac{1}{3}\mu_B + \frac{2}{3}\mu_Q,\\
\mu_d &= \frac{1}{3}\mu_B - \frac{1}{3}\mu_Q,\\
\mu_s &= \frac{1}{3}\mu_B - \frac{1}{3}\mu_Q - \mu_S.
\label{eq:muqg}
\end{align}
Equation~\eqref{eq:lnz_gluon} involves one-dimensional momentum integrals over the semi-infinite domain $p \in [0, \infty)$. To facilitate backpropagation through this component of the DNN, we implement a 50-point Gauss-quadrature method for numerical integration. For the sake of brevity, the explicit dependence of thermodynamic quantities on $(T, \mu_B, \mu_Q, \mu_S)$ will be omitted in the following expressions.

Then, the corresponding thermodynamic quantities, such as pressure $p$, energy density $\varepsilon$, entropy density $s$, particle number density $n_X$ (where $X$ represents $B$, $Q$ or $S$) and trace anomaly $\Delta$, are obtained using automatic differentiation techniques, with the specific calculation formulas as follows:
\begin{align}
p &= T \left( \frac{\partial \ln Z}{\partial V} \right)_{T,\mu_B,\mu_Q,\mu_S}, \\
s &=  \left( \frac{\partial p}{\partial T} \right)_{\mu_B,\mu_Q,\mu_S}, \\
n_B &=\left( \frac{\partial p}{\partial \mu_B} \right)_{T,\mu_Q,\mu_S}, \\
n_Q &=\left( \frac{\partial p}{\partial \mu_Q} \right)_{T,\mu_B,\mu_S}, \\
n_S &=\left( \frac{\partial p}{\partial \mu_S} \right)_{T,\mu_B,\mu_Q}, \\
\varepsilon &= Ts  - p + \sum_X^{B,Q,S}\mu_X n_X, \\ 
\Delta &=\frac{\varepsilon  - 3 P}{T^4}.
\label{eq:eos}
\end{align}

In this study, we incorporate the most recent generalized susceptibilities $\chi_{i,j,k}^{B,Q,S}$ derived from LQCD calculations\cite{Abuali:2025tbd} to guarantee that our quasi-parton masses accurately reflect the properties of QCD matter across the $\mu_Q$ and $\mu_S$ dimensions. The $\chi_{i,j,k}^{B,Q,S}$ is given by:
\begin{align}
\chi_{i,j,k}^{B,Q,S} &= \frac{\partial{p/T^4}}{\partial{\hat{\mu}_B^i}\partial{\hat{\mu}_Q^j}\partial{\hat{\mu}_S^k}}
\bigg |_{\hat{\mu}_B=\hat{\mu}_Q=\hat{\mu}_S= 0}, \nonumber \\ 
\hat{\mu}_B& = \mu_B/T, \hat{\mu}_Q = \mu_Q/T, \hat{\mu}_S = \mu_S/T,
\label{eq:chib_def}
\end{align}
the loss function consists of five distinct components: 
\begin{align}
    \mathcal{L}(\theta_1, \theta_2, \theta_3)&=\left|s_\text {{DNN}}-s_{\text {LQCD }}\right|+\left|\frac{ \Delta_\text{{DNN}}-\Delta_{\text {LQCD}}}{T}\right| \\
    & +\left|\chi_{i,j,k, \text {DNN}}^{B,Q,S}- \chi_{i,j,k, \text {LQCD}}^{B,Q,S}\right|+\left|n_{X,\text{DNN}}\big|_{\mu_X=0}\right| \nonumber\\
    &+\mathcal{L}_\text {{MC}} \nonumber
\end{align}
The $\theta_1$, $\theta_2$ and $\theta_3$ are parameters in the three mass model. The $n_X$ is the particle number density, which should be zero when $\mu_B=\mu_Q=\mu_S=0$. Additionally, the mass constraint term $\mathcal{L}_{MC}$ is derived from the HTL, as demonstrated in \cite{Haque:2024gva,Levai:1997yx}:
\begin{align}
 \mathcal{L}_\text{MC} & = \mathcal{L}_1 + \mathcal{L}_2 , \nonumber \\
 &= \left| R_{g/q} - \frac{3}{2} \right| +  \left| \frac{m_s - m_{u/d}}{\overline{m}_s - \overline{m}_{u/d}} -1\right|, \nonumber \\
R_{g/q} &= \frac{M_{g,T>2.5T_\text{{cut}}}}{M_{q,T>2.5T_\text{{cut}}}} = \sqrt{\frac{3}{2}\left(\frac{N_C}{3} + \frac{N_f}{6}\right)},
\label{eq:MCC}
\end{align} 
where $T_\text{{cut}}$ =0.150 GeV, $\overline{m}_s$ and $\overline{m}_{u/d}$ are current mass. Finally, the loss function is optimized based on the gradient descent algorithm, which effectively determines the optimal parameter configuration ( $\theta_1$, $\theta_2$ and $\theta_3$) in the mass model.

The training data are sourced from lattice simulations \cite{Abuali:2025tbd}. We uniformly selected 70 temperature points within the range of 0.05 to 0.4 GeV at $\mu_B = 0$. Subsequently, the trained model is tested at finite $\mu_B$\cite{Abuali:2025tbd}. We do not perform a full uncertainty propagation in this work because the lattice data employ continuum estimates with correlated systematic errors whose full covariance matrix is not publicly available. Independent sampling of the input data might induce anomalous behavior in higher-order $\chi_{i,j,k}^{B,Q,S}$, as their correlations are essential for maintaining thermodynamic consistency. A rigorous Bayesian or ensemble uncertainty analysis will be pursued in future work. 

\section{Results and discussion}
\label{Res}

\begin{figure}[htbp]
\centering
\includegraphics[width=0.48\textwidth]{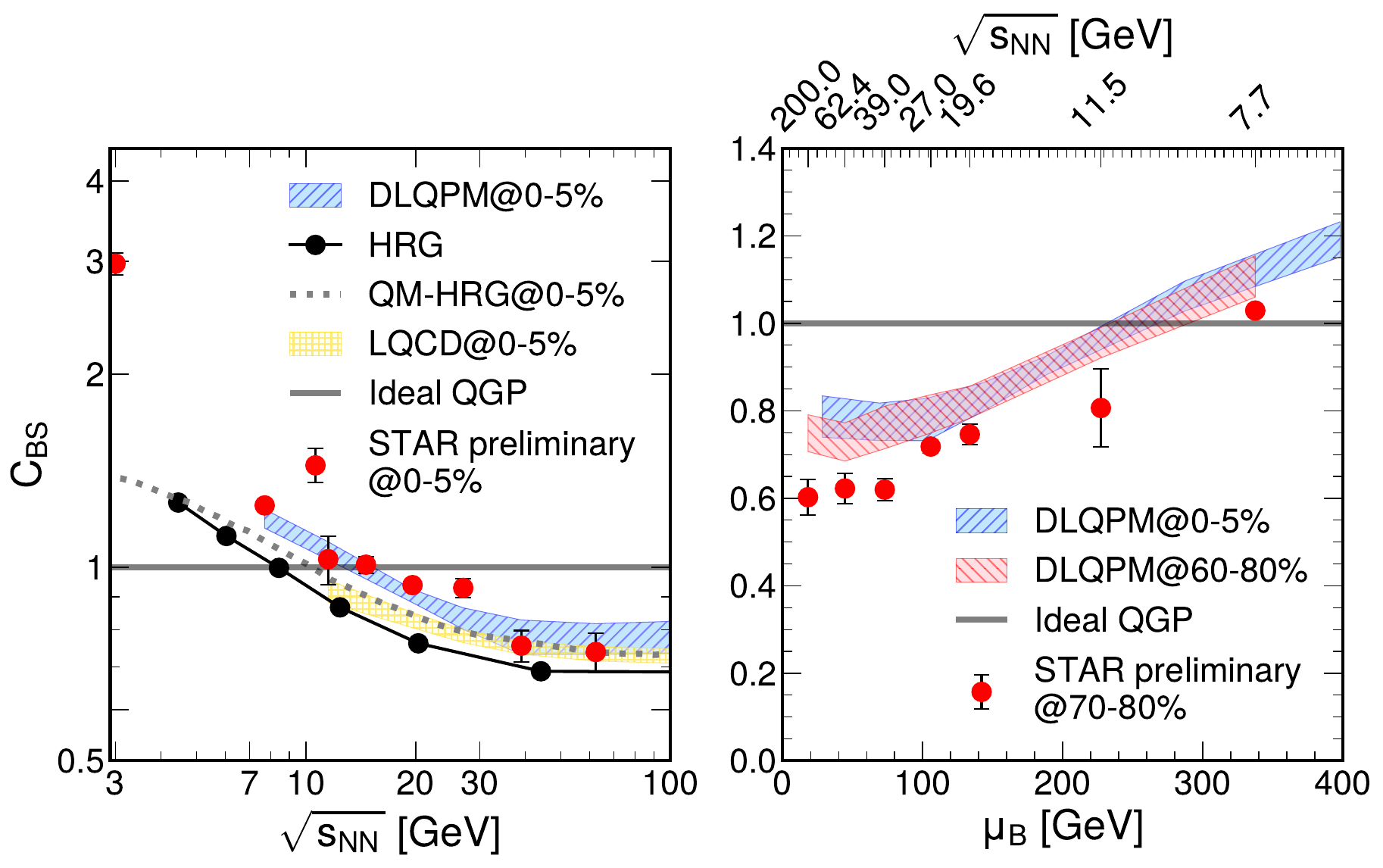}
\caption {\label{fig:CBS}(Color online) The baryon-strangeness correlation coefficient $C_{BS}$ is presented as functions of $\sqrt{s_{NN}}$ and $\mu_B$ for different centrality classes. The left panel displays $C_{BS}$ as a function of $\sqrt{s_{NN}}$ for 0-5\% centrality, with comparisons to the LQCD calculations, HRG model, QM-HRG model and preliminary STAR results \cite{Bollweg:2024epj,Bollweg:2021vqf,STAR-QM}. The red data points correspond to STAR experimental measurements, the yellow band denotes LQCD results, the black dotted line represents HRG model outcomes, and the dashed line indicates QM-HRG model predictions. The color bands show the results from DLQPM. The right panel presents $C_{BS}$ as a function of baryon chemical potential ($\mu_B$) for two centrality classes: 0-5\% (most central) and 60-80\% (peripheral, shown in blue and pink).}
\end{figure}

\begin{figure*}[htbp]
\centering
\includegraphics[width=0.98\textwidth]{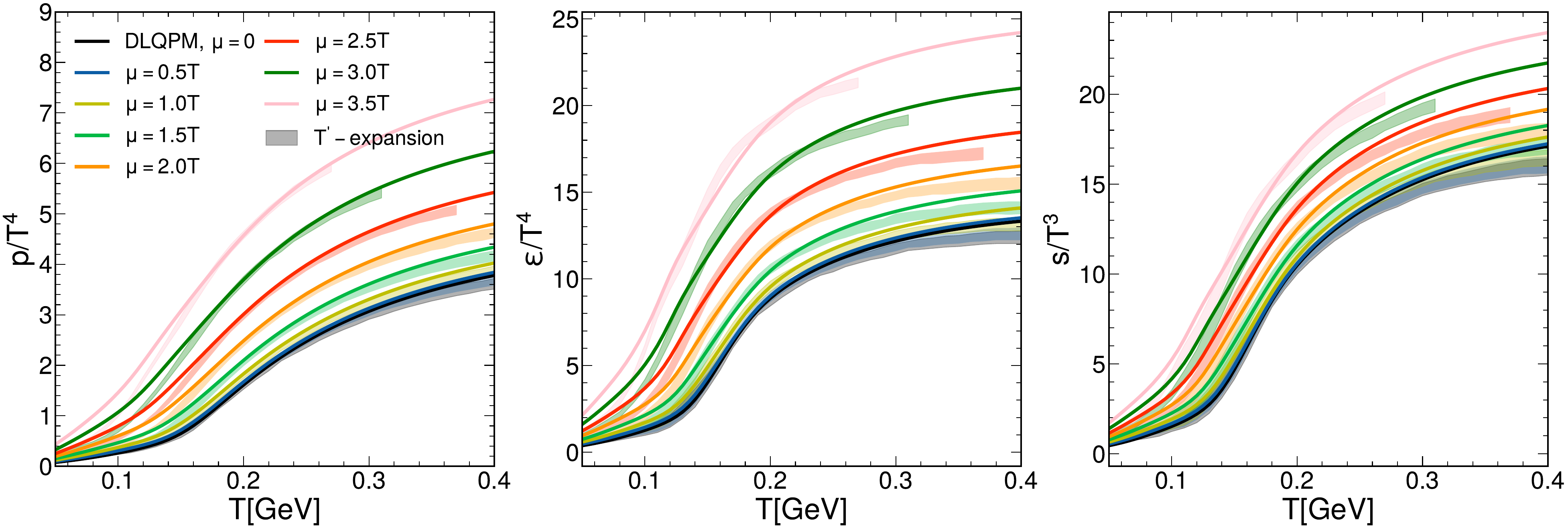}
\caption {\label{fig:QCDEOS}(Color online) The pressure $p/T^4$, energy density $\varepsilon/T^4$ and entropy density $s/T^3$ as functions of $T$. The solid lines represent the computational results of our model, while the bands denote the calculation results from $T^{\prime}$-expansion\cite{Abuali:2025tbd}, with different colors corresponding to distinct values of $\mu/T$.}
\end{figure*}

Fig~\ref{fig:CBS} shows the baryon–strangeness correlation coefficient $C_{BS}$ as a function of the beam energy $\sqrt{s_{NN}}$, compared with lattice QCD (LQCD), the hadron resonance gas (HRG) model, the quark-model–inspired HRG (QM-HRG) model, and preliminary STAR data \cite{Bollweg:2024epj,Bollweg:2021vqf,STAR-QM}. The QM-HRG model incorporates additional hadronic states predicted by the quark model \cite{Bazavov:2014xya}. As shown in the left panel, all theoretical calculations exhibit a monotonic decrease of $C_{BS}$ with increasing $\sqrt{s_{NN}}$. The DLQPM results are in good agreement with the preliminary STAR data in the energy range $\sqrt{s_{NN}}$ = 7.7–100~GeV for 0–5\% centrality. Calculations at $\sqrt{s_{NN}} = 3$~GeV are not shown, as the corresponding freeze-out parameters $(T_{ch}, \mu_B, \mu_Q, \mu_S)$ are available only for beam energies above 7.7~GeV \cite{STAR:2017sal}. Overall, the DLQPM predictions are consistent with LQCD results, although some slight discrepancies exist at lower beam energies. Within uncertainties, our calculations are compatible with the preliminary STAR measurements.

The right panel of fig.~\ref{fig:CBS} presents $C_{BS}$ as a function of the baryon chemical potential $\mu_B$ for peripheral collisions. Since the freeze-out parameters $(T_{ch}, \mu_B, \mu_Q, \mu_S)$ are available only for the 60–80\% centrality class, the DLQPM results are evaluated in this centrality range and compared with preliminary STAR data for 70–80\% centrality. The DLQPM calculations reproduce the overall trend observed in the experimental data but yield values that are approximately 10\% higher, particularly at low $\mu_B$. The model also exhibits a negligible dependence on collision centrality. This deviation may be partly attributed to the use of freeze-out parameters extracted from statistical thermal models, which assume local thermal equilibrium—an assumption that may be less justified in peripheral collisions. Nevertheless, these results highlight the sensitivity of $C_{BS}$ to the underlying equation of state. Future measurements of $C_{BS}$ over a broad range of $\sqrt{s_{NN}}$ and collision centralities will provide valuable constraints for refining the DLQPM and improving the precision of the reconstructed four-dimensional QCD EoS.

Fig.~\ref{fig:QCDEOS} presents $p/T^4$, $\varepsilon/T^4$ and $s/T^3$ as functions of $T$ at various $\mu/T$ ranging from 0 to 3.5, with comparison to $T^\prime$-expansion calculations \cite{Abuali:2025tbd}. The chemical potential $\mu$ is defined as $\mu=\sqrt{\mu_B^2 + \mu_Q^2 + \mu_S^2}$, under the constraint that $\mu_B/\sqrt{2}$= $\mu_Q$ = $\mu_S$. The DLQPM-calculated QCD EoS agrees with the $T^\prime$-expansion results in both temperature-dependent behavior and numerical values, particularly above 0.1 GeV. Below this temperature threshold, the DLQPM yields slightly higher values compared to the $T^\prime$-expansion method. This deviation is most pronounced at the lowest temperatures shown and should be kept in mind when applying the EoS in the hadronic regime.

\begin{figure*}[htbp]
\centering
\includegraphics[width=0.98\textwidth]{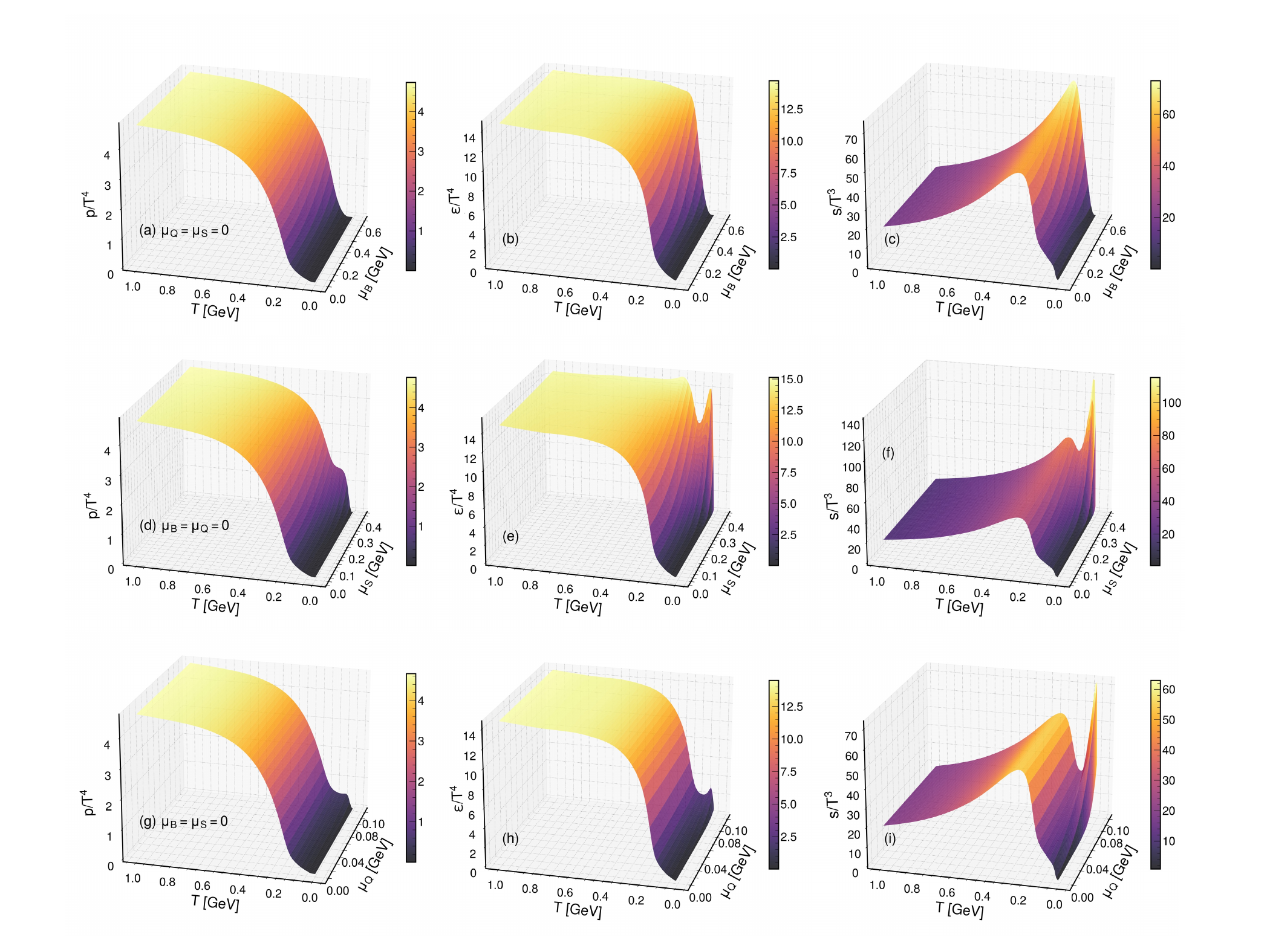}
\caption {\label{fig:EoSall}(Color online) The $p/T^4$, $\epsilon/T^4$ and $s/T^3$ across the $T$ - $\mu_B$, $T$ - $\mu_S$ and $T$ - $\mu_Q$ planes.}
\end{figure*}

Fig.~\ref{fig:EoSall} displays the DLQPM results for $p/T^4$, $\epsilon/T^4$, and $s/T^3$ across the two-dimensional planes of ($T$, $\mu_B$), ($T$, $\mu_S$), and ($T$, $\mu_Q$). The figure shows that as the chemical-potential-to-temperature ratio increases, the EoS gradually diverges from the $\mu=0$ baseline; this trend is most pronounced on the ($T$, $\mu_S$) and ($T$, $\mu_Q$) planes. Because the DLQPM is currently constrained only by lattice data at $\mu_B=0$, the extrapolation to the low-temperature, high-density region carries larger systematic uncertainties. In future work, we will utilize available experimental data, including centrality-dependent $C_{BS}$ measurements, to better constrain the EoS in this regime.

\section{Summary and outlook}
\label{Sum}

In this study, we systematically investigate the 4D QCD EoS by incorporating temperature- and chemical potential-dependent quasi-parton masses—specifically, for baryon ($\mu_B$), electric charge ($\mu_Q$), and strangeness ($\mu_S$) chemical potentials—within a thermodynamic framework. Our results demonstrate that the 4D DLQPM accurately reproduces the QCD EoS from LQCD at $\mu_B=0$, as well as the susceptibilities $\chi_{i,j,k}^{B,Q,S}$ at vanishing chemical potentials $\mu_X = 0$, where $X = B, Q, S$. Furthermore, the EoS derived under various chemical potential conditions using the DLQPM shows excellent agreement with results obtained via the $T^\prime$-expansion method.

Additionally, we compute the baryon-strangeness correlation coefficient $C_{BS}$ as a function of the center-of-mass energy per nucleon pair $\sqrt{s_{NN}}$ and compare our results with the latest experimental data, LQCD predictions, the QM-HRG model, and the UrQMD model. Our analysis reveals that the DLQPM yields good agreement with the preliminary STAR data in central collisions. Discrepancies observed in peripheral collisions suggest limitations in the current model, possibly due to non-equilibrium effects not fully captured by the DLQPM or to the use of freeze-out parameters extracted from equilibrium statistical models.

Looking ahead, we plan to enhance the model’s representational capacity by integrating experimental data—particularly centrality-dependent $C_{BS}$ measurements and EoS constraints near the QCD critical point—as optimization targets. We also intend to develop the DLQPM into an accessible software package to support hydrodynamic simulations by providing direct access to the 4D QCD EoS and transport coefficients such as the shear viscosity $\eta$ and bulk viscosity $\xi$. Moreover, future work will apply the DLQPM to systematically explore the properties of QCD matter under strong magnetic fields \cite{Ding:2021cwv,Kumar:2025ikm}. These developments are expected to significantly improve the reliability and adaptability of the framework for studying QCD thermodynamics and transport properties.

\section{acknowledgments}

The authors thank Akihiko Monnai for providing the NEoS data. This work was supported by the National Natural Science Foundation of China (NSFC) under Grant Nos.\ 12075098, 12435009, 12225503 and in part by the China Postdoctoral Science Foundation under Grant No.\ 2025M783370. Financial support was also provided by the self-determined research funds of Central China Normal University (CCNU) from the colleges' basic research and operation of the Ministry of Education (MOE). We gratefully acknowledge the extensive computing resources provided by the Nuclear Computing Center of Central China Normal University.

For further information about the data and code used in this study, please contact the authors via email.

\section*{Declaration of generative AI and AI-assisted technologies in the writing process}

During the preparation of this work the author(s) used DeepSeek in order to polish the language. After using this tool/service, the author(s) reviewed and edited the content as needed and take(s) full responsibility for the content of the publication.

\bibliography{ref}

@article{Monnai:2019hkn,
    author = {Monnai, Akihiko and Schenke, Bj\"orn and Shen, Chun},
    title = "{Equation of state at finite densities for QCD matter in nuclear collisions}",
    eprint = "1902.05095",
    archivePrefix = "arXiv",
    primaryClass = "nucl-th",
    reportNumber = "KEK-TH-2106",
    doi = "10.1103/PhysRevC.100.024907",
    journal = "Phys. Rev. C",
    volume = "100",
    number = "2",
    pages = "024907",
    year = "2019"
}

@article{An:2021wof,
    author = "An, Xin and others",
    title = "{The BEST framework for the search for the QCD critical point and the chiral magnetic effect}",
    eprint = "2108.13867",
    archivePrefix = "arXiv",
    primaryClass = "nucl-th",
    doi = "10.1016/j.nuclphysa.2021.122343",
    journal = "Nucl. Phys. A",
    volume = "1017",
    pages = "122343",
    year = "2022"
}

@article{Denicol:2018wdp,
    author = {Denicol, Gabriel S. and Gale, Charles and Jeon, Sangyong and Monnai, Akihiko and Schenke, Bj\"orn and Shen, Chun},
    title = "{Net baryon diffusion in fluid dynamic simulations of relativistic heavy-ion collisions}",
    eprint = "1804.10557",
    archivePrefix = "arXiv",
    primaryClass = "nucl-th",
    reportNumber = "KEK-TH-2075",
    doi = "10.1103/PhysRevC.98.034916",
    journal = "Phys. Rev. C",
    volume = "98",
    number = "3",
    pages = "034916",
    year = "2018"
}

@article{MUSES:2023hyz,
    author = "Kumar, Rajesh and others",
    collaboration = "MUSES",
    title = "{Theoretical and experimental constraints for the equation of state of dense and hot matter}",
    eprint = "2303.17021",
    archivePrefix = "arXiv",
    primaryClass = "nucl-th",
    doi = "10.1007/s41114-024-00049-6",
    journal = "Living Rev. Rel.",
    volume = "27",
    number = "1",
    pages = "3",
    year = "2024"
}

@article{Borsanyi:2022qlh,
    author = "Borsanyi, Szabolcs and Guenther, Jana N. and Kara, Ruben and Fodor, Zoltan and Parotto, Paolo and Pasztor, Attila and Ratti, Claudia and Szabo, Kalman K.",
    title = "{Resummed lattice QCD equation of state at finite baryon density: Strangeness neutrality and beyond}",
    eprint = "2202.05574",
    archivePrefix = "arXiv",
    primaryClass = "hep-lat",
    doi = "10.1103/PhysRevD.105.114504",
    journal = "Phys. Rev. D",
    volume = "105",
    number = "11",
    pages = "114504",
    year = "2022"
}

@article{HotQCD:2014kol,
    author = "Bazavov, A. and others",
    collaboration = "HotQCD",
    title = "{Equation of state in ( 2+1 )-flavor QCD}",
    eprint = "1407.6387",
    archivePrefix = "arXiv",
    primaryClass = "hep-lat",
    reportNumber = "BNL-105928-2014-JA",
    doi = "10.1103/PhysRevD.90.094503",
    journal = "Phys. Rev. D",
    volume = "90",
    pages = "094503",
    year = "2014"
}

@article{Monnai:2021kgu,
    author = {Monnai, Akihiko and Schenke, Bj\"orn and Shen, Chun},
    title = "{QCD Equation of State at Finite Chemical Potentials for Relativistic Nuclear Collisions}",
    eprint = "2101.11591",
    archivePrefix = "arXiv",
    primaryClass = "nucl-th",
    doi = "10.1142/S0217751X21300076",
    journal = "Int. J. Mod. Phys. A",
    volume = "36",
    number = "07",
    pages = "2130007",
    year = "2021"
}

@article{Levai:1997yx,
    author = "Levai, Peter and Heinz, Ulrich W.",
    title = "{Massive gluons and quarks and the equation of state obtained from SU(3) lattice QCD}",
    eprint = "hep-ph/9710463",
    archivePrefix = "arXiv",
    doi = "10.1103/PhysRevC.57.1879",
    journal = "Phys. Rev. C",
    volume = "57",
    pages = "1879--1890",
    year = "1998"
}

@article{Shi:2021qri,
    author = "Shi, Shuzhe and Zhou, Kai and Zhao, Jiaxing and Mukherjee, Swagato and Zhuang, Pengfei",
    title = "{Heavy quark potential in the quark-gluon plasma: Deep neural network meets lattice quantum chromodynamics}",
    eprint = "2105.07862",
    archivePrefix = "arXiv",
    primaryClass = "hep-ph",
    doi = "10.1103/PhysRevD.105.014017",
    journal = "Phys. Rev. D",
    volume = "105",
    number = "1",
    pages = "014017",
}

@article{Alqahtani:2015qja,
    author = "Alqahtani, Mubarak and Nopoush, Mohammad and Strickland, Michael",
    title = "{Quasiparticle equation of state for anisotropic hydrodynamics}",
    eprint = "1509.02913",
    archivePrefix = "arXiv",
    primaryClass = "hep-ph",
    doi = "10.1103/PhysRevC.92.054910",
    journal = "Phys. Rev. C",
    volume = "92",
    number = "5",
    pages = "054910",
    year = "2015"
}

@article{Liu:2021dpm,
    author = "Liu, Feng-Lei and Xing, Wen-Jing and Wu, Xiang-Yu and Qin, Guang-You and Cao, Shanshan and Wang, Xin-Nian",
    title = "{QLBT: a linear Boltzmann transport model for heavy quarks in a quark-gluon plasma of quasi-particles}",
    eprint = "2107.11713",
    archivePrefix = "arXiv",
    primaryClass = "hep-ph",
    doi = "10.1140/epjc/s10052-022-10308-x",
    journal = "Eur. Phys. J. C",
    volume = "82",
    number = "4",
    pages = "350",
    year = "2022"
}

@phdthesis{2653187,
    author = "Mykhaylova, Valeriya",
    title = "{Transport Properties of Hot QCD Matter in the Quasiparticle Approach}"
}

@article{Mykhaylova:2020pfk,
    author = "Mykhaylova, Valeriya and Sasaki, Chihiro",
    title = "{Impact of quark quasiparticles on transport coefficients in hot QCD}",
    eprint = "2007.06846",
    archivePrefix = "arXiv",
    primaryClass = "hep-ph",
    doi = "10.1103/PhysRevD.103.014007",
    journal = "Phys. Rev. D",
    volume = "103",
    number = "1",
    pages = "014007",
    year = "2021"
}

@article{Kang:2022jbg,
    author = "Kang, Zhaofeng and Zhu, Jiang and Guo, Jun",
    title = "{Massive gauge theory with quasigluon for hot SU(N): Phase transition and thermodynamics}",
    eprint = "2211.09442",
    archivePrefix = "arXiv",
    primaryClass = "hep-ph",
    doi = "10.1103/PhysRevD.107.076005",
    journal = "Phys. Rev. D",
    volume = "107",
    number = "7",
    pages = "076005",
    year = "2023"
}

@article{RAISSI2019686,
    author = "M. Raissi, P. Perdikaris and G.E. Karniadakis",
    title = "{Physics-informed neural networks: A deep learning framework for solving forward and inverse problems involving nonlinear partial differential equations}",
    doi = "https://doi.org/10.1016/j.jcp.2018.10.045",
    journal = "Journal of Computational Physics",
    volume = "378",
    pages = "686-707",
    year = "2019"
}

@article{Plumari:2011mk,
    author = "Plumari, Salvatore and Alberico, Wanda M. and Greco, Vincenzo and Ratti, Claudia",
    title = "{Recent thermodynamic results from lattice QCD analyzed within a quasi-particle model}",
    eprint = "1103.5611",
    archivePrefix = "arXiv",
    primaryClass = "hep-ph",
    doi = "10.1103/PhysRevD.84.094004",
    journal = "Phys. Rev. D",
    volume = "84",
    pages = "094004",
    year = "2011"
}

@article{Bernhard:2019bmu,
    author = "Bernhard, Jonah E. and Moreland, J. Scott and Bass, Steffen A.",
    title = "{Bayesian estimation of the specific shear and bulk viscosity of quark-gluon plasma}",
    doi = "10.1038/s41567-019-0611-8",
    journal = "Nature Phys.",
    volume = "15",
    number = "11",
    pages = "1113--1117",
    year = "2019"
}

@article{Peshier:1995ty,
    author = "Peshier, A. and Kampfer, Burkhard and Pavlenko, O. P. and Soff, G.",
    title = "{A Massive quasiparticle model of the SU(3) gluon plasma}",
    reportNumber = "PRINT-96-141 (DRESDEN)",
    doi = "10.1103/PhysRevD.54.2399",
    journal = "Phys. Rev. D",
    volume = "54",
    pages = "2399--2402",
    year = "1996"
}

@article{Koch:2025cog,
    author = "Koch, Volker and Vovchenko, Volodymyr",
    title = "{Exploring the QCD phase diagram through correlations and fluctuations}",
    eprint = "2512.04288",
    archivePrefix = "arXiv",
    primaryClass = "nucl-th",
    month = "12",
    year = "2025"
}

@article{Borsanyi:2013bia,
    author = "Borsanyi, Szabocls and Fodor, Zoltan and Hoelbling, Christian and Katz, Sandor D. and Krieg, Stefan and Szabo, Kalman K.",
    title = "{Full result for the QCD equation of state with 2+1 flavors}",
    eprint = "1309.5258",
    archivePrefix = "arXiv",
    primaryClass = "hep-lat",
    doi = "10.1016/j.physletb.2014.01.007",
    journal = "Phys. Lett. B",
    volume = "730",
    pages = "99--104",
    year = "2014"
}

@article{Hornik1989MultilayerFN,
  title={Multilayer feedforward networks are universal approximators},
  author={Kurt Hornik and Maxwell B. Stinchcombe and Halbert L. White},
  journal={Neural Networks},
  year={1989},
  volume={2},
  pages={359-366}
}

@article{Soma:2023fyi,
    author = {Soma, Shriya and Wang, Lingxiao and Shi, Shuzhe and St\"ocker, Horst and Zhou, Kai},
    title = "{A physics-based neural network reconstruction of the dense matter equation of state from neutron star observables}",
    doi = "10.1051/epjconf/202327606007",
    journal = "EPJ Web Conf.",
    volume = "276",
    pages = "06007",
    year = "2023"
}

@article{Wang:2024dzc,
    author = "Wang, Lingxiao",
    title = "{Deep learning for exploring hadron\textendash{}hadron interactions}",
    eprint = "2501.00374",
    archivePrefix = "arXiv",
    primaryClass = "nucl-th",
    reportNumber = "RIKEN-iTHEMS-Report-24",
    doi = "10.1016/j.jspc.2025.100024",
    journal = "J. Subatomic Part. Cosmol.",
    volume = "3",
    pages = "100024",
    year = "2025"
}

@article{2018PINN,
  title={Physics-Informed Neural Networks: A Deep Learning Framework for Solving Forward and Inverse Problems Involving Nonlinear Partial Differential Equations},
  author={Raissi, M.  and  Perdikaris, P.  and  Karniadakis, G. E. },
  journal={Journal of Computational Physics},
  year={2019},
  volume={378},
  pages={686-707},
  doi={https://doi.org/10.1016/j.jcp.2018.10.045}
}

@article{JMLR:v18:17-468,
  author  = {Atilim Gunes Baydin and Barak A. Pearlmutter and Alexey Andreyevich Radul and Jeffrey Mark Siskind},
  title   = {Automatic Differentiation in Machine Learning: a Survey},
  journal = {Journal of Machine Learning Research},
  year    = {2018},
  volume  = {18},
  number  = {153},
  pages   = {1-43},
  url     = {http://jmlr.org/papers/v18/17-468.html}
}

@article{Pang:2016vdc,
    author = {Pang, Long-Gang and Zhou, Kai and Su, Nan and Petersen, Hannah and St\"ocker, Horst and Wang, Xin-Nian},
    title = "{An equation-of-state-meter of quantum chromodynamics transition from deep learning}",
    eprint = "1612.04262",
    archivePrefix = "arXiv",
    primaryClass = "hep-ph",
    doi = "10.1038/s41467-017-02726-3",
    journal = "Nature Commun.",
    volume = "9",
    number = "1",
    pages = "210",
    year = "2018"
}

@article{1976A,
  title={A GUIDE TO FEYNMAN DIAGRAMS IN THE MANY-BODY PROBLEM},
  author={ Mattuck, R. },
  journal={Osborne McGraw-Hill},
  year={1976},
}

@article{Zhou:2023pti,
    author = "Zhou, Kai and Wang, Lingxiao and Pang, Long-Gang and Shi, Shuzhe",
    title = "{Exploring QCD matter in extreme conditions with Machine Learning}",
    eprint = "2303.15136",
    archivePrefix = "arXiv",
    primaryClass = "hep-ph",
    doi = "10.1016/j.ppnp.2023.104084",
    journal = "Prog. Part. Nucl. Phys.",
    volume = "135",
    pages = "104084",
    year = "2024"
}

@article{Pang:2012he,
    author = "Pang, Longgang and Wang, Qun and Wang, Xin-Nian",
    title = "{Effects of initial flow velocity fluctuation in event-by-event (3+1)D hydrodynamics}",
    eprint = "1205.5019",
    archivePrefix = "arXiv",
    primaryClass = "nucl-th",
    reportNumber = "NT-LBL-12-011",
    doi = "10.1103/PhysRevC.86.024911",
    journal = "Phys. Rev. C",
    volume = "86",
    pages = "024911",
    year = "2012"
}

@article{Shen:2014vra,
    author = "Shen, Chun and Qiu, Zhi and Song, Huichao and Bernhard, Jonah and Bass, Steffen and Heinz, Ulrich",
    title = "{The iEBE-VISHNU code package for relativistic heavy-ion collisions}",
    eprint = "1409.8164",
    archivePrefix = "arXiv",
    primaryClass = "nucl-th",
    doi = "10.1016/j.cpc.2015.08.039",
    journal = "Comput. Phys. Commun.",
    volume = "199",
    pages = "61--85",
    year = "2016"
}

@article{Pang:2018zzo,
    author = "Pang, Long-Gang and Petersen, Hannah and Wang, Xin-Nian",
    title = "{Pseudorapidity distribution and decorrelation of anisotropic flow within the open-computing-language implementation CLVisc hydrodynamics}",
    eprint = "1802.04449",
    archivePrefix = "arXiv",
    primaryClass = "nucl-th",
    doi = "10.1103/PhysRevC.97.064918",
    journal = "Phys. Rev. C",
    volume = "97",
    number = "6",
    pages = "064918",
    year = "2018"
}

@article{Achenbach:2023pba,
    author = "Achenbach, P. and others",
    title = "{The present and future of QCD}",
    eprint = "2303.02579",
    archivePrefix = "arXiv",
    primaryClass = "hep-ph",
    reportNumber = "JLAB-PHY-23-3808",
    doi = "10.1016/j.nuclphysa.2024.122874",
    journal = "Nucl. Phys. A",
    volume = "1047",
    pages = "122874",
    year = "2024"
}

@book{thuerey2021pbdl,
  title={Physics-based Deep Learning},
  author={Nils Thuerey and Philipp Holl and Maximilian Mueller and Patrick Schnell and Felix Trost and Kiwon Um},
  url={https://physicsbaseddeeplearning.org},
  year={2021},
  publisher={WWW}
}

@article{Soloveva:2021quj,
    author = {Soloveva, Olga and Aichelin, J\"org and Bratkovskaya, Elena},
    title = "{Transport properties and equation-of-state of hot and dense QGP matter near the critical endpoint in the phenomenological dynamical quasiparticle model}",
    eprint = "2108.08561",
    archivePrefix = "arXiv",
    primaryClass = "hep-ph",
    doi = "10.1103/PhysRevD.105.054011",
    journal = "Phys. Rev. D",
    volume = "105",
    number = "5",
    pages = "054011",
    year = "2022"
}

@misc{ramachandran2017searchingactivationfunctions,
      title={Searching for Activation Functions}, 
      author={Prajit Ramachandran and Barret Zoph and Quoc V. Le},
      year={2017},
      eprint={1710.05941},
      archivePrefix={arXiv},
      primaryClass={cs.NE},
      url={https://arxiv.org/abs/1710.05941}, 
}

@article{Ding:2021cwv,
    author = "Ding, H. -T. and Li, S. -T. and Shi, Q. and Wang, X. -D.",
    title = "{Fluctuations and correlations of net baryon number, electric charge and strangeness in a background magnetic field}",
    eprint = "2104.06843",
    archivePrefix = "arXiv",
    primaryClass = "hep-lat",
    doi = "10.1140/epja/s10050-021-00519-3",
    journal = "Eur. Phys. J. A",
    volume = "57",
    number = "6",
    pages = "202",
    year = "2021"
}

@article{Kumar:2025ikm,
    author = "Kumar, Arpith and Ding, Heng-Tong and Gu, Jin-Biao and Li, Sheng-Tai",
    title = "{QCD Equation of State with Strong Magnetic Fields and Nonzero Baryon Density}",
    eprint = "2502.03152",
    archivePrefix = "arXiv",
    primaryClass = "hep-lat",
    doi = "10.22323/1.466.0175",
    journal = "PoS",
    volume = "LATTICE2024",
    pages = "175",
    year = "2025"
}

@article{Luo:2017faz,
    author = "Luo, Xiaofeng and Xu, Nu",
    title = "{Search for the QCD Critical Point with Fluctuations of Conserved Quantities in Relativistic Heavy-Ion Collisions at RHIC : An Overview}",
    eprint = "1701.02105",
    archivePrefix = "arXiv",
    primaryClass = "nucl-ex",
    doi = "10.1007/s41365-017-0257-0",
    journal = "Nucl. Sci. Tech.",
    volume = "28",
    number = "8",
    pages = "112",
    year = "2017"
}

@article{Li:2022ozl,
    author = {Li, Fu-Peng and L\"u, Hong-Liang and Pang, Long-Gang and Qin, Guang-You},
    title = "{Deep-learning quasi-particle masses from QCD equation of state}",
    eprint = "2211.07994",
    archivePrefix = "arXiv",
    primaryClass = "hep-ph",
    doi = "10.1016/j.physletb.2023.138088",
    journal = "Phys. Lett. B",
    volume = "844",
    pages = "138088",
    year = "2023"
}

@article{Bollweg:2022rps,
    author = "Bollweg, D. and Goswami, J. and Kaczmarek, O. and Karsch, F. and Mukherjee, Swagato and Petreczky, P. and Schmidt, C. and Scior, P.",
    collaboration = "HotQCD",
    title = "{Taylor expansions and Pad\'e approximants for cumulants of conserved charge fluctuations at nonvanishing chemical potentials}",
    eprint = "2202.09184",
    archivePrefix = "arXiv",
    primaryClass = "hep-lat",
    doi = "10.1103/PhysRevD.105.074511",
    journal = "Phys. Rev. D",
    volume = "105",
    number = "7",
    pages = "074511",
    year = "2022"
}

@article{Elfner:2022iae,
    author = {Elfner, Hannah and M\"uller, Berndt},
    title = "{The exploration of hot and dense nuclear matter: introduction to relativistic heavy-ion physics}",
    eprint = "2210.12056",
    archivePrefix = "arXiv",
    primaryClass = "nucl-th",
    doi = "10.1088/1361-6471/ace824",
    journal = "J. Phys. G",
    volume = "50",
    number = "10",
    pages = "103001",
    year = "2023"
}

@article{HotQCD:2018pds,
    author = "Bazavov, A. and others",
    collaboration = "HotQCD",
    title = "{Chiral crossover in QCD at zero and non-zero chemical potentials}",
    eprint = "1812.08235",
    archivePrefix = "arXiv",
    primaryClass = "hep-lat",
    doi = "10.1016/j.physletb.2019.05.013",
    journal = "Phys. Lett. B",
    volume = "795",
    pages = "15--21",
    year = "2019"
}

@article{Monnai:2024pvy,
    author = {Monnai, Akihiko and Pihan, Gr\'egoire and Schenke, Bj\"orn and Shen, Chun},
    title = "{Four-dimensional QCD equation~of state with multiple chemical potentials}",
    eprint = "2406.11610",
    archivePrefix = "arXiv",
    primaryClass = "nucl-th",
    doi = "10.1103/PhysRevC.110.044905",
    journal = "Phys. Rev. C",
    volume = "110",
    number = "4",
    pages = "044905",
    year = "2024"
}

@article{HotQCD:2012fhj,
    author = "Bazavov, A. and others",
    collaboration = "HotQCD",
    title = "{Fluctuations and Correlations of net baryon number, electric charge, and strangeness: A comparison of lattice QCD results with the hadron resonance gas model}",
    eprint = "1203.0784",
    archivePrefix = "arXiv",
    primaryClass = "hep-lat",
    doi = "10.1103/PhysRevD.86.034509",
    journal = "Phys. Rev. D",
    volume = "86",
    pages = "034509",
    year = "2012"
}

@book{Kapusta:2006pm,
    author = "Kapusta, J. I. and Gale, Charles",
    title = "{Finite-temperature field theory: Principles and applications}",
    doi = "10.1017/CBO9780511535130",
    isbn = "978-0-521-17322-3, 978-0-521-82082-0, 978-0-511-22280-1",
    publisher = "Cambridge University Press",
    series = "Cambridge Monographs on Mathematical Physics",
    year = "2011"
}

@article{Haque:2024gva,
    author = "Haque, Najmul and Mustafa, Munshi G.",
    title = "{Hard Thermal Loop\textemdash{}Theory and applications}",
    eprint = "2404.08734",
    archivePrefix = "arXiv",
    primaryClass = "hep-ph",
    doi = "10.1016/j.ppnp.2024.104136",
    journal = "Prog. Part. Nucl. Phys.",
    volume = "140",
    pages = "104136",
    year = "2025"
}

@article{Braun-Munzinger:2015hba,
    author = {Braun-Munzinger, Peter and Koch, Volker and Sch\"afer, Thomas and Stachel, Johanna},
    title = "{Properties of hot and dense matter from relativistic heavy ion collisions}",
    eprint = "1510.00442",
    archivePrefix = "arXiv",
    primaryClass = "nucl-th",
    doi = "10.1016/j.physrep.2015.12.003",
    journal = "Phys. Rept.",
    volume = "621",
    pages = "76--126",
    year = "2016"
}

@article{Sorensen:2023zkk,
    author = "Sorensen, Agnieszka and others",
    title = "{Dense nuclear matter equation of state from heavy-ion collisions}",
    eprint = "2301.13253",
    archivePrefix = "arXiv",
    primaryClass = "nucl-th",
    reportNumber = "INT-PUB-23-001, LA-UR-23-20514, LLNL-TR-844629",
    doi = "10.1016/j.ppnp.2023.104080",
    journal = "Prog. Part. Nucl. Phys.",
    volume = "134",
    pages = "104080",
    year = "2024"
}

@article{Aarts2025,
  author    = {Gert Aarts and Kenji Fukushima and Tetsuo Hatsuda and Andreas Ipp and Shuzhe Shi and Lingxiao Wang and Kai Zhou},
  title     = {Physics-driven learning for inverse problems in quantum chromodynamics},
  journal   = {Nature Reviews Physics},
  year      = {2025},
  date      = {2025-01-06},
  volume    = {7},
  number    = {1},
  pages     = {1--15},
  issn      = {2522-5820},
  doi       = {10.1038/s42254-024-00798-x},
  url       = {https://doi.org/10.1038/s42254-024-00798-x}
}

@article{Soloveva:2023tvj,
    author = "Soloveva, Olga and Palermo, Andrea and Bratkovskaya, Elena",
    title = "{Extraction of the microscopic properties of quasiparticles using deep neural networks}",
    eprint = "2311.15984",
    archivePrefix = "arXiv",
    primaryClass = "hep-ph",
    doi = "10.1103/PhysRevC.110.034908",
    journal = "Phys. Rev. C",
    volume = "110",
    number = "3",
    pages = "034908",
    year = "2024"
}

@article{Pang:2024kid,
    author = "Pang, Long-Gang",
    title = "{Studying high-energy nuclear physics with machine learning}",
    doi = "10.1142/S0218301324300091",
    journal = "Int. J. Mod. Phys. E",
    volume = "33",
    number = "06",
    pages = "2430009",
    year = "2024"
}

@inproceedings{NEURIPS2019_bdbca288,
    author = {Paszke, Adam and Gross, Sam and Massa, Francisco and Lerer, Adam and Bradbury, James and Chanan, Gregory and Killeen, Trevor and Lin, Zeming and Gimelshein, Natalia and Antiga, Luca and Desmaison, Alban and Kopf, Andreas and Yang, Edward and DeVito, Zachary and Raison, Martin and Tejani, Alykhan and Chilamkurthy, Sasank and Steiner, Benoit and Fang, Lu and Bai, Junjie and Chintala, Soumith},
    booktitle = {Advances in Neural Information Processing Systems},
    editor = {H. Wallach and H. Larochelle and A. Beygelzimer and F. d\textquotesingle Alch\'{e}-Buc and E. Fox and R. Garnett},
    pages = {},
    publisher = {Curran Associates, Inc.},
    title = {PyTorch: An Imperative Style, High-Performance Deep Learning Library},
    url = {https://proceedings.neurips.cc/paper_files/paper/2019/file/bdbca288fee7f92f2bfa9f7012727740-Paper.pdf},
    volume = {32},
    year = {2019}
}

@article{Romatschke:2011qp,
    author = "Romatschke, Paul",
    title = "{Relativistic (Lattice) Boltzmann Equation with Non-Ideal Equation of State}",
    eprint = "1108.5561",
    archivePrefix = "arXiv",
    primaryClass = "gr-qc",
    doi = "10.1103/PhysRevD.85.065012",
    journal = "Phys. Rev. D",
    volume = "85",
    pages = "065012",
    year = "2012"
}

@article{Bollweg:2024epj,
    author = "Bollweg, D. and Ding, H. -T. and Goswami, J. and Karsch, F. and Mukherjee, Swagato and Petreczky, P. and Schmidt, C.",
    title = "{Strangeness-correlations on the pseudocritical line in (2+1)-flavor QCD}",
    eprint = "2407.09335",
    archivePrefix = "arXiv",
    primaryClass = "hep-lat",
    doi = "10.1103/PhysRevD.110.054519",
    journal = "Phys. Rev. D",
    volume = "110",
    number = "5",
    pages = "054519",
    year = "2024"
}

@article{Borsanyi:2021sxv,
    author = "Bors\'anyi, S. and Fodor, Z. and Guenther, J. N. and Kara, R. and Katz, S. D. and Parotto, P. and P\'asztor, A. and Ratti, C. and Szab\'o, K. K.",
    title = "{Lattice QCD equation of state at finite chemical potential from an alternative expansion scheme}",
    eprint = "2102.06660",
    archivePrefix = "arXiv",
    primaryClass = "hep-lat",
    doi = "10.1103/PhysRevLett.126.232001",
    journal = "Phys. Rev. Lett.",
    volume = "126",
    number = "23",
    pages = "232001",
    year = "2021"
}

@article{Mondal:2021jxk,
    author = "Mondal, Sourav and Mukherjee, Swagato and Hegde, Prasad",
    title = "{Lattice QCD Equation of State for Nonvanishing Chemical Potential by Resumming Taylor Expansions}",
    eprint = "2106.03165",
    archivePrefix = "arXiv",
    primaryClass = "hep-lat",
    doi = "10.1103/PhysRevLett.128.022001",
    journal = "Phys. Rev. Lett.",
    volume = "128",
    number = "2",
    pages = "022001",
    year = "2022"
}

@article{Bollweg:2021vqf,
    author = "Bollweg, D. and Goswami, J. and Kaczmarek, O. and Karsch, F. and Mukherjee, Swagato and Petreczky, P. and Schmidt, C. and Scior, P.",
    collaboration = "HotQCD",
    title = "{Second order cumulants of conserved charge fluctuations revisited: Vanishing chemical potentials}",
    eprint = "2107.10011",
    archivePrefix = "arXiv",
    primaryClass = "hep-lat",
    doi = "10.1103/PhysRevD.104.074512",
    journal = "Phys. Rev. D",
    volume = "104",
    number = "7",
    year = "2021"
}

@misc{STAR-QM,
    author = "Feng, Hanwen",
    collaboration = "for the STAR Collaboration",
        title = "{Baryon-Strangeness Correlations in Au+Au Collisions at RHIC-STAR, XXXI International Conference on Ultra-relativistic Nucleus-Nucleus Collisions, Frankfurt, Apr.6-12, 2025}",
        doi = "https://drupal.star.bnl.gov/STAR/files/hwfeng_qm2025.pdf",
}

@article{Koch:2006ek,
    author = "Koch, V. and Majumder, A. and Randrup, J.",
    editor = "Csorgo, T. and Levai, P. and David, G. and Papp, G.",
    title = "{Baryon-strangeness correlations: a diagnostic of strongly interacting matter}",
    doi = "10.1016/j.nuclphysa.2006.06.147",
    journal = "Nucl. Phys. A",
    volume = "774",
    pages = "841--844",
    year = "2006"
}

@article{Koch:2005vg,
    author = "Koch, V. and Majumder, A. and Randrup, J.",
    title = "{Baryon-strangeness correlations: A Diagnostic of strongly interacting matter}",
    eprint = "nucl-th/0505052",
    archivePrefix = "arXiv",
    reportNumber = "LBNL-57613",
    doi = "10.1103/PhysRevLett.95.182301",
    journal = "Phys. Rev. Lett.",
    volume = "95",
    pages = "182301",
    year = "2005"
}

@article{Wu:2021fjf,
    author = "Wu, Xiang-Yu and Qin, Guang-You and Pang, Long-Gang and Wang, Xin-Nian",
    title = "{(3+1)-D viscous hydrodynamics at finite net baryon density: Identified particle spectra, anisotropic flows, and flow fluctuations across energies relevant to the beam-energy scan at RHIC}",
    eprint = "2107.04949",
    archivePrefix = "arXiv",
    primaryClass = "hep-ph",
    doi = "10.1103/PhysRevC.105.034909",
    journal = "Phys. Rev. C",
    volume = "105",
    number = "3",
    pages = "034909",
    year = "2022"
}

@article{Abuali:2025tbd,
    author = "Abuali, Ahmed and Bors{\'a}nyi, Szabolcs and Fodor, Zolt{\'a}n and Jahan, Johannes and Kahangirwe, Micheal and Parotto, Paolo and P{\'a}sztor, Attila and Ratti, Claudia and Shah, Hitansh and Trabulsi, Seth A.",
    title = "{New 4D lattice QCD equation of state: Extended density coverage from a generalized T' expansion}",
    eprint = "2504.01881",
    archivePrefix = "arXiv",
    primaryClass = "hep-lat",
    doi = "10.1103/2dmh-26yh",
    journal = "Phys. Rev. D",
    volume = "112",
    number = "5",
    pages = "054502",
    year = "2025"
}

@article{Li:2025csc,
    author = "Li, Fu-Peng and Pang, Long-Gang and Qin, Guang-You",
    title = "{QCD equation of state at finite {\ensuremath{\mu}}B using deep learning assisted quasi-parton model}",
    eprint = "2501.10012",
    archivePrefix = "arXiv",
    primaryClass = "nucl-th",
    doi = "10.1016/j.physletb.2025.139692",
    journal = "Phys. Lett. B",
    volume = "868",
    pages = "139692",
    year = "2025"
}

@article{STAR:2017sal,
    author = "Adamczyk, L. and others",
    collaboration = "STAR",
    title = "{Bulk Properties of the Medium Produced in Relativistic Heavy-Ion Collisions from the Beam Energy Scan Program}",
    eprint = "1701.07065",
    archivePrefix = "arXiv",
    primaryClass = "nucl-ex",
    doi = "10.1103/PhysRevC.96.044904",
    journal = "Phys. Rev. C",
    volume = "96",
    number = "4",
    pages = "044904",
    year = "2017"
}

@article{Cheng:2008zh,
    author = "Cheng, M. and others",
    title = "{Baryon Number, Strangeness and Electric Charge Fluctuations in QCD at High Temperature}",
    eprint = "0811.1006",
    archivePrefix = "arXiv",
    primaryClass = "hep-lat",
    reportNumber = "BI-TP-2008-35",
    doi = "10.1103/PhysRevD.79.074505",
    journal = "Phys. Rev. D",
    volume = "79",
    pages = "074505",
    year = "2009"
}

@article{Bazavov:2013dta,
    author = "Bazavov, A. and others",
    title = "{Strangeness at high temperatures: from hadrons to quarks}",
    eprint = "1304.7220",
    archivePrefix = "arXiv",
    primaryClass = "hep-lat",
    doi = "10.1103/PhysRevLett.111.082301",
    journal = "Phys. Rev. Lett.",
    volume = "111",
    pages = "082301",
    year = "2013"
}

@inbook{Koch:2008ia,
    author = "Koch, Volker",
    editor = "Stock, R.",
    title = "{Hadronic Fluctuations and Correlations}",
    booktitle = "{Relativistic Heavy Ion Physics}",
    eprint = "0810.2520",
    archivePrefix = "arXiv",
    primaryClass = "nucl-th",
    doi = "10.1007/978-3-642-01539-7_20",
    pages = "626--652",
    year = "2010"
}

\appendix
\section{The training results}

\begin{figure}[htbp]
\centering
\includegraphics[width=0.48\textwidth]{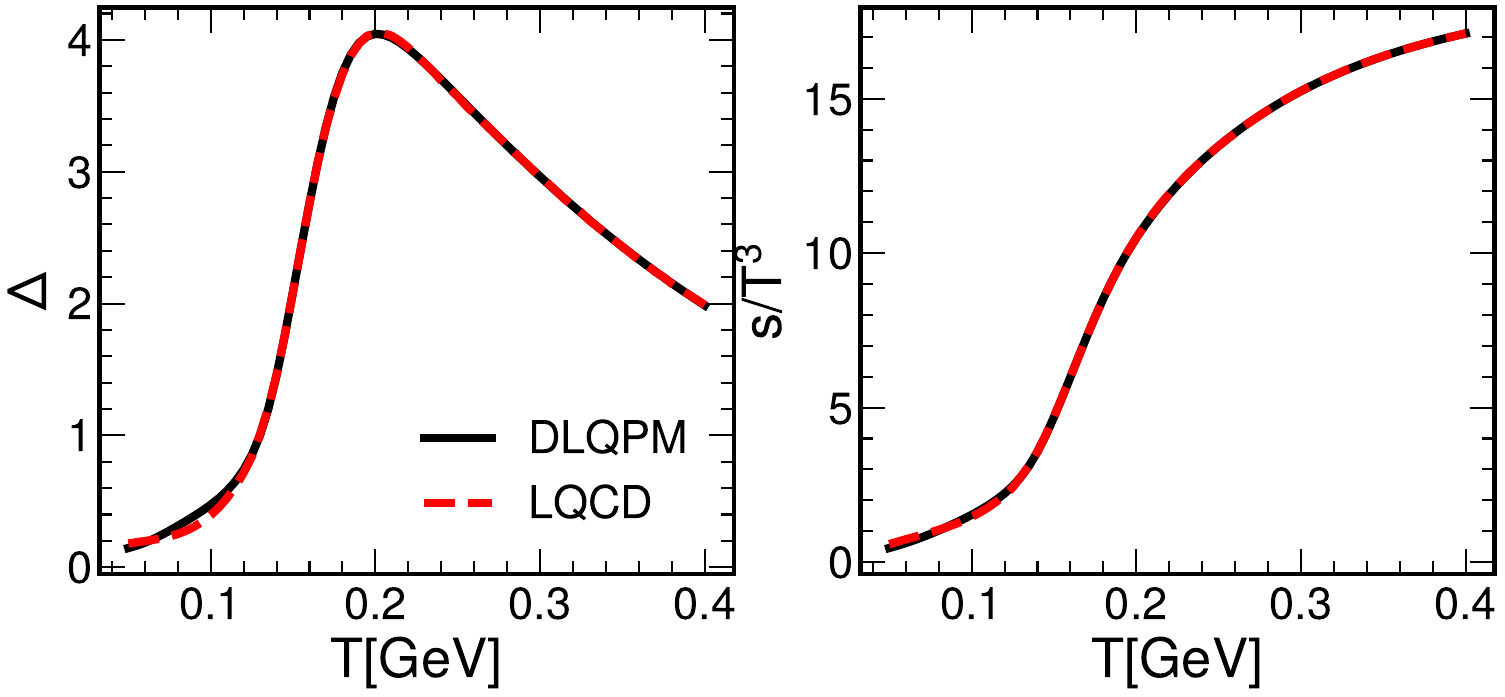}
\caption {\label{fig:s_delta}(Color online) The  trace anomaly $\Delta$ and entropy density $s/T^3$ as a function of temperature. The red dashed curve represents LQCD calculations\cite{HotQCD:2014kol,Abuali:2025tbd}, while the black solid curve corresponds to results obtained from the DLQPM. }
\end{figure}

Fig.~\ref{fig:s_delta} illustrates the temperature dependence of both the trace anomaly $\Delta$ and the entropy density $s/T^3$ at $\mu_B=0$. The results indicate that the DLQPM can accurately reproduce the QCD EoS. As temperature increases, the trace anomaly $\Delta$ first rises to a maximum value before decreasing, whereas the scaled entropy density $s/T^3$ shows a continuous monotonic increase with temperature.

\begin{figure*}[htbp]
\centering
\includegraphics[width=0.98\textwidth]{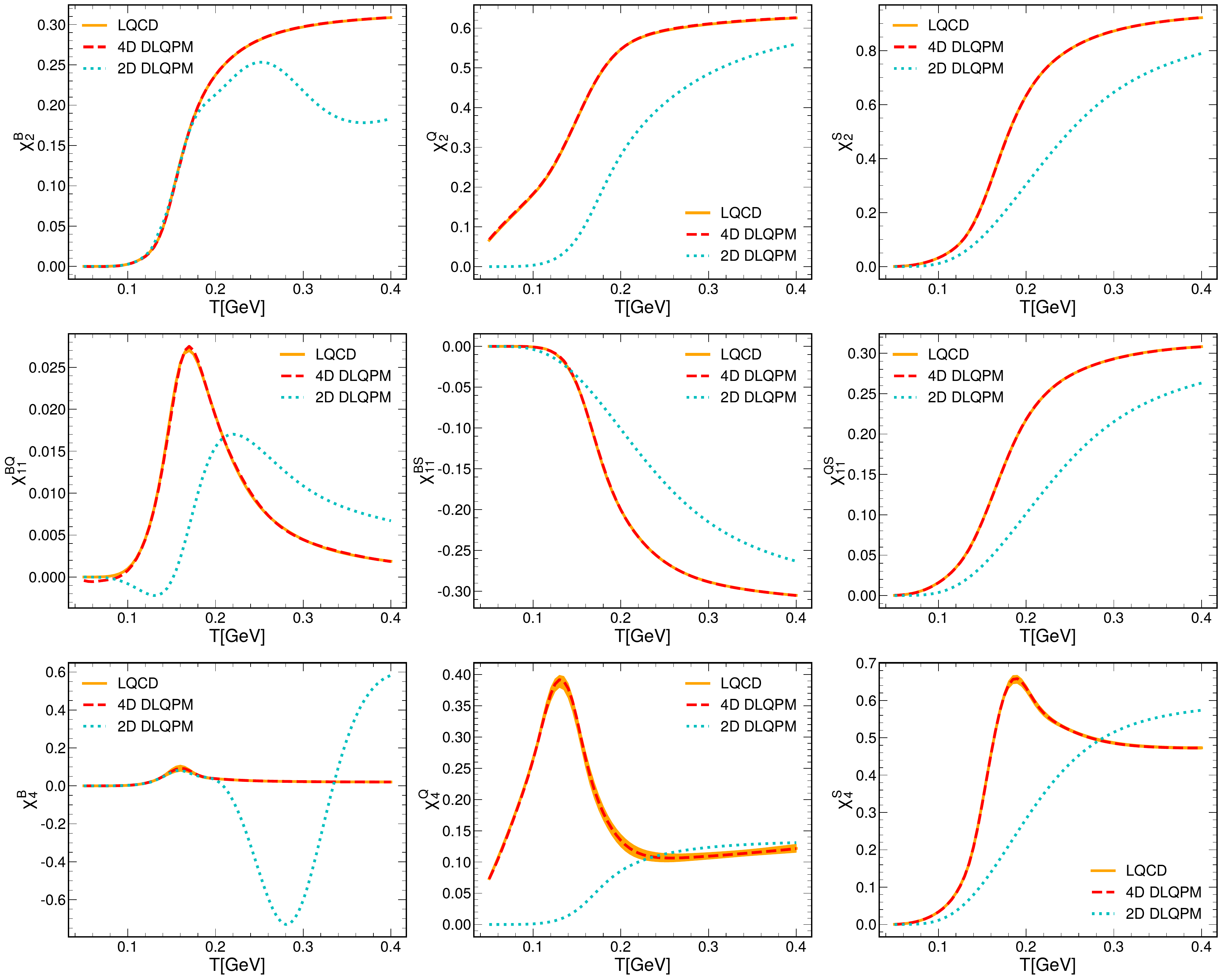}
\caption {\label{fig:chi_BQS1}(Color online) The temperature-dependent susceptibilities 
$\chi_{i,j,k}^{B,Q,S}$ are compared with LQCD results and 2D DLQPM calculations \cite{Abuali:2025tbd,Li:2025csc}. The comparison includes three distinct calculations: (1) LQCD results shown as orange band, (2) 4D DLQPM calculations represented by a red dashed line, and (3) 2D DLQPM calculations indicated by a cyan dotted line.}
\end{figure*}

\begin{figure*}[htbp]
\centering
\includegraphics[width=0.98\textwidth]{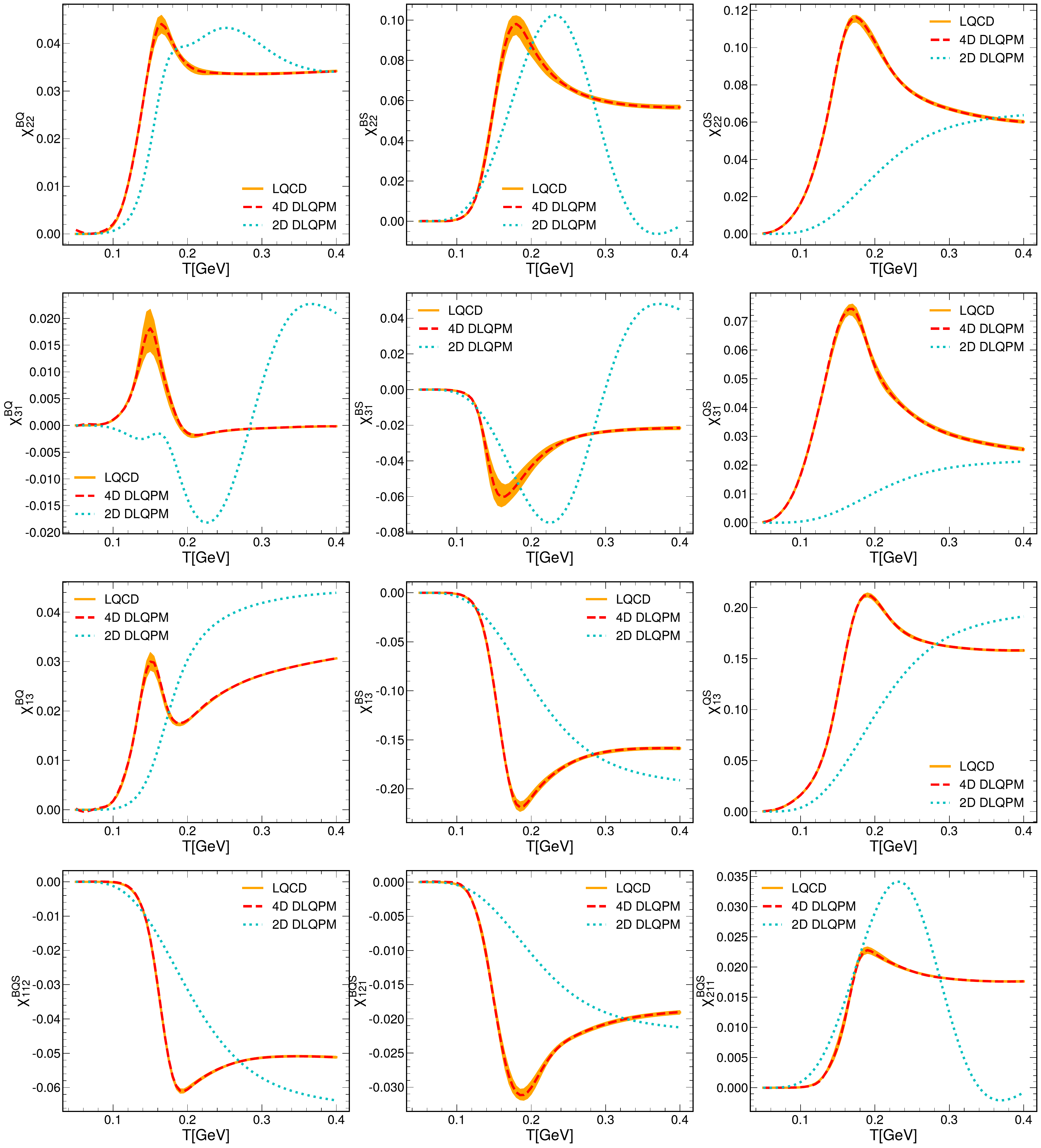}
\caption {\label{fig:chi_BQS2}(Color online) Same as in fig.~\ref{fig:chi_BQS1}.}
\end{figure*}

Here, we conducted a comparative analysis of the results for $\chi_{i,j,k}^{B,Q,S}$. The Figs.~\ref{fig:chi_BQS1} and \ref{fig:chi_BQS2} present the temperature dependence of $\chi_{i,j,k}^{B,Q,S}$ in comparison with LQCD data and the previous DLQPM (2D DLQPM) results  \cite{Li:2025csc}.  The current DLQPM calculations demonstrate remarkable consistency with LQCD predictions. Unlike the 2D DLQPM which accounted solely for $T$ and $\mu_B$ dependencies, the present implementation integrates additional physical constraints, thereby achieving enhanced representation in 4D space.

\begin{figure*}
  \centering
  \includegraphics[width=0.98\linewidth]{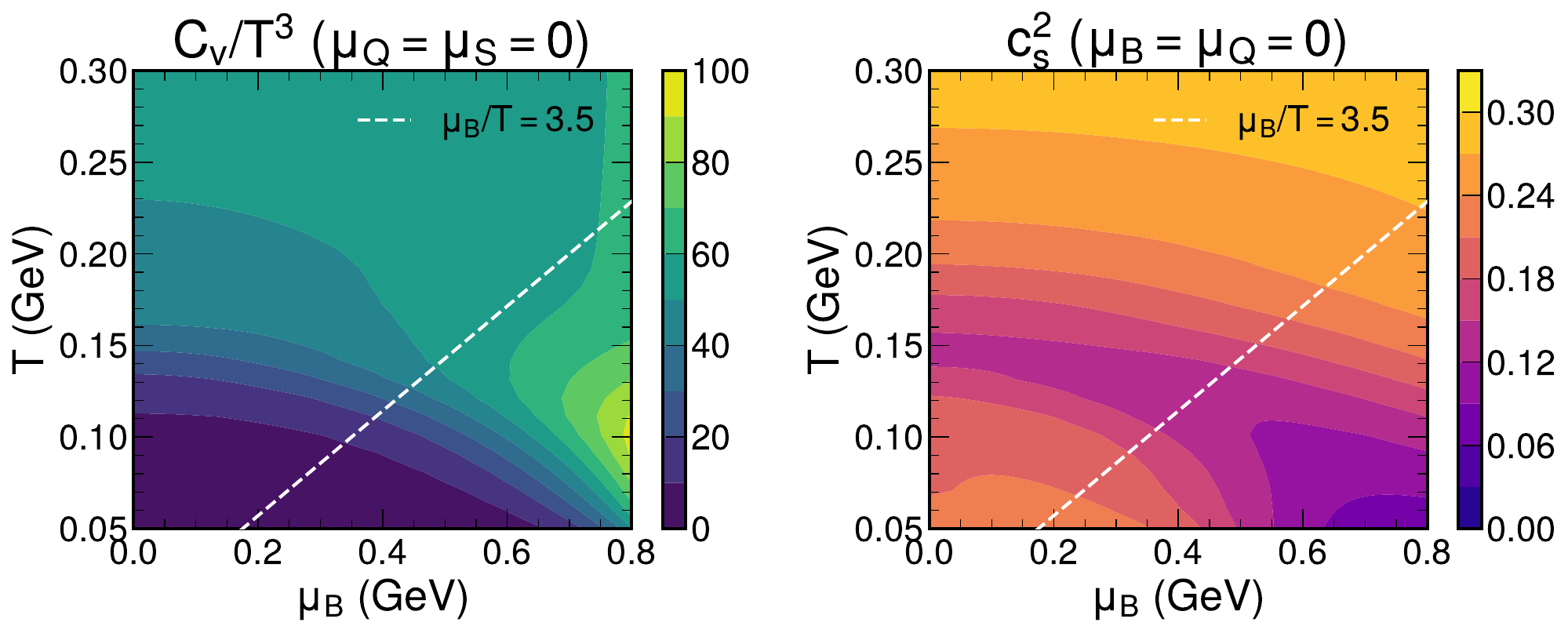} \\
  \vspace{1em}      
  \includegraphics[width=0.98\linewidth]{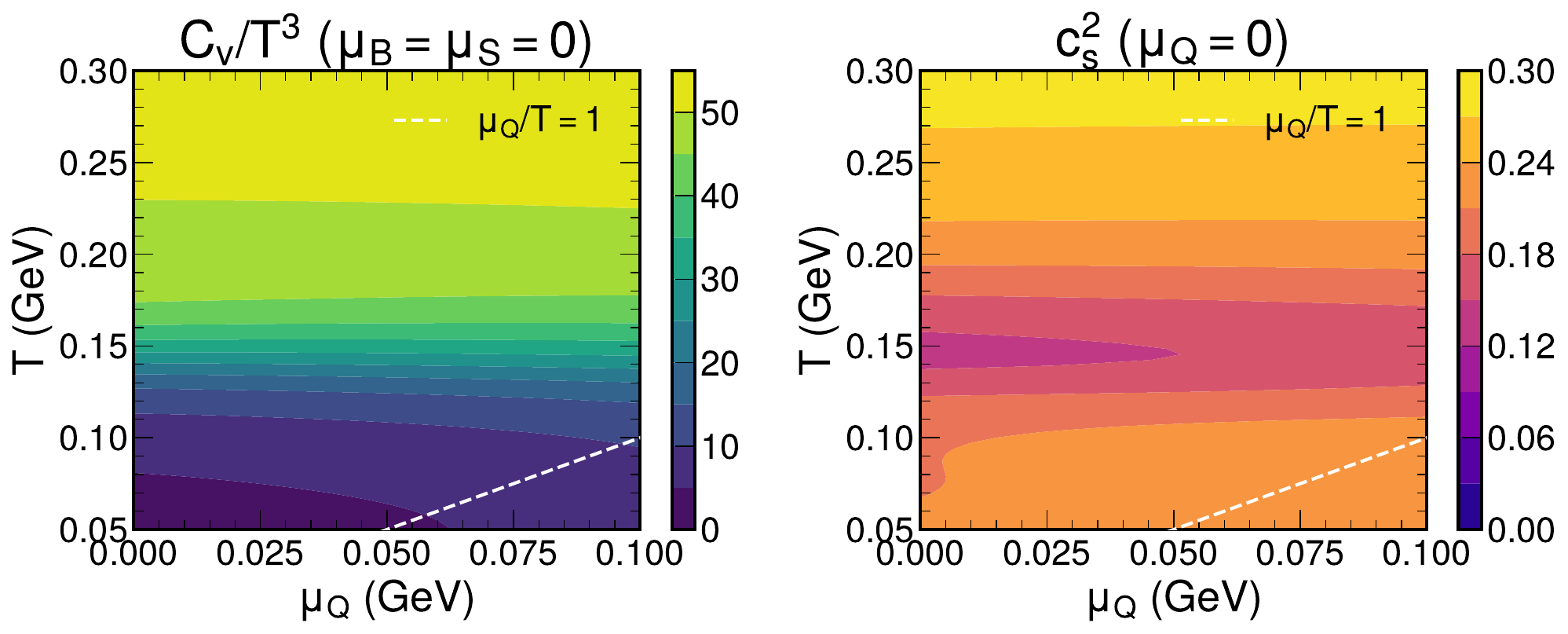} \\
  \vspace{1em}
  \includegraphics[width=0.98\linewidth]{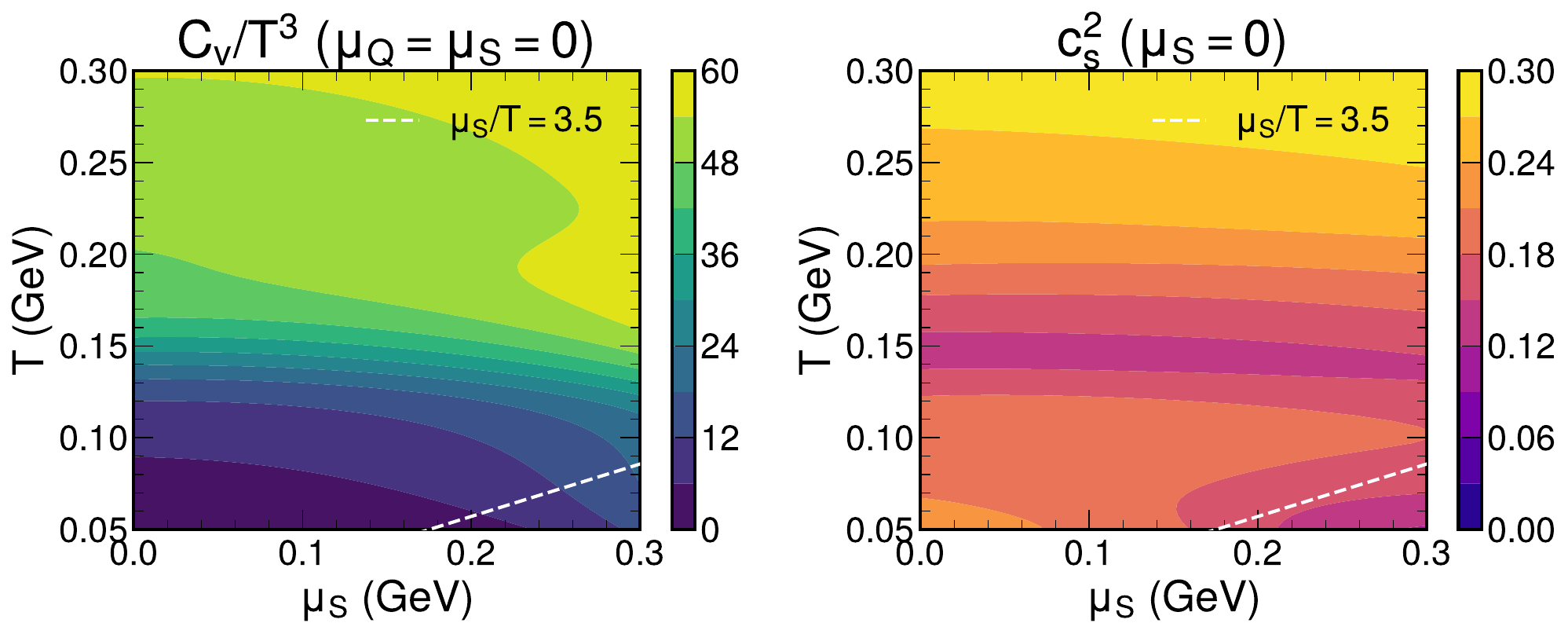}
  \caption{(Color online) The specific heat $C_V$ (left panels) and squared speed of sound $c_s^2$ (right panels) as functions of temperature $T$ for different chemical potential configurations. Top panels: $\mu_Q=\mu_S=0$; Middle panels: $\mu_B=\mu_S=0$; Bottom panels: $\mu_B=\mu_Q=0$. }
  \label{fig:Cv_cs}
\end{figure*}

Fig.~\ref{fig:Cv_cs} shows the specific heat $C_V$ and speed of sound $c_s^2$ as functions of temperature $T$ for different chemical potential configurations.

\begin{figure*}
  \centering
  \includegraphics[width=0.98\linewidth]{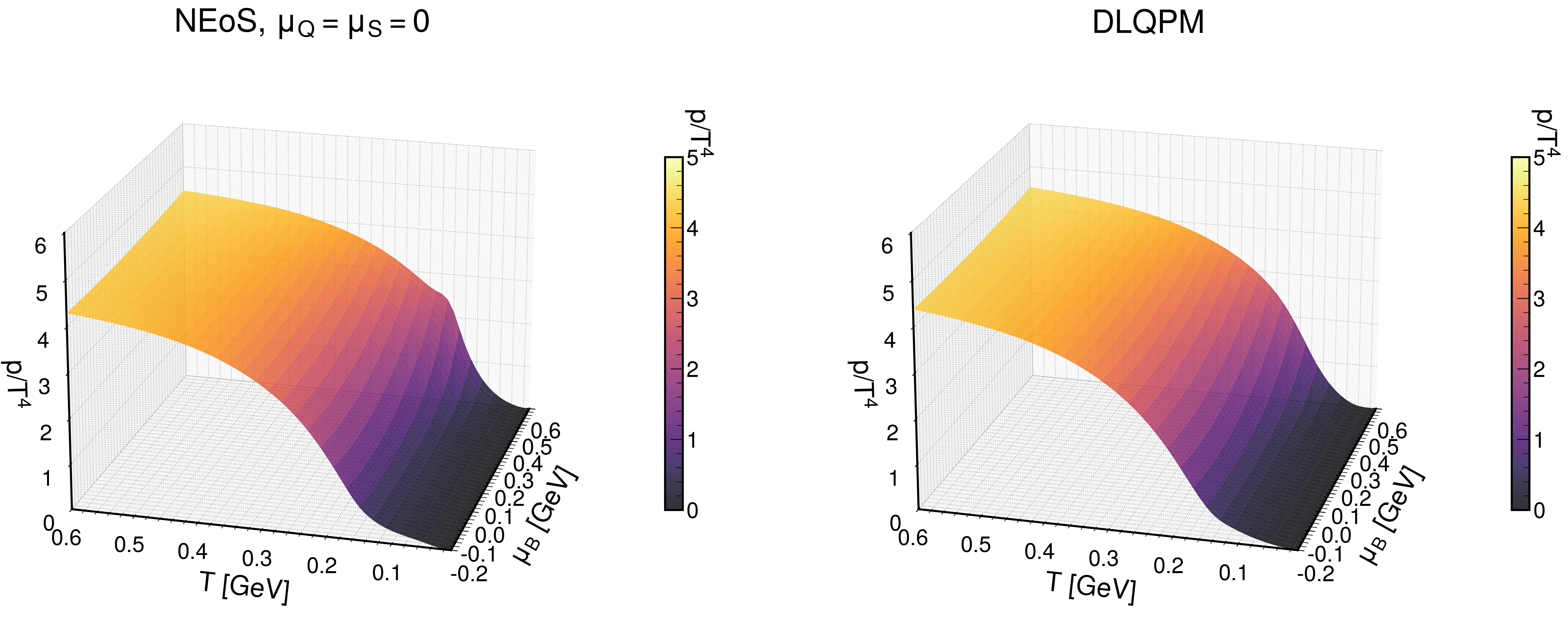} \\
  \vspace{1em}      
  \includegraphics[width=0.98\linewidth]{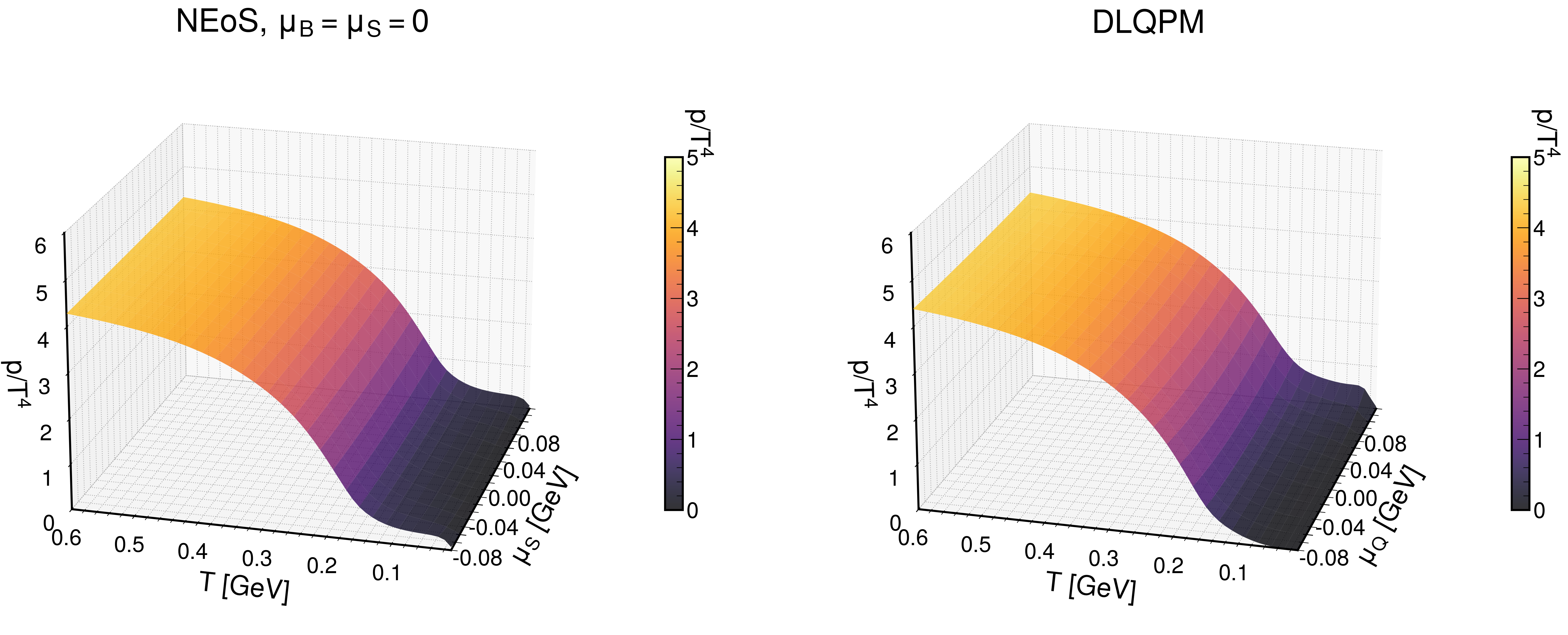} \\
  \vspace{1em}
  \includegraphics[width=0.98\linewidth]{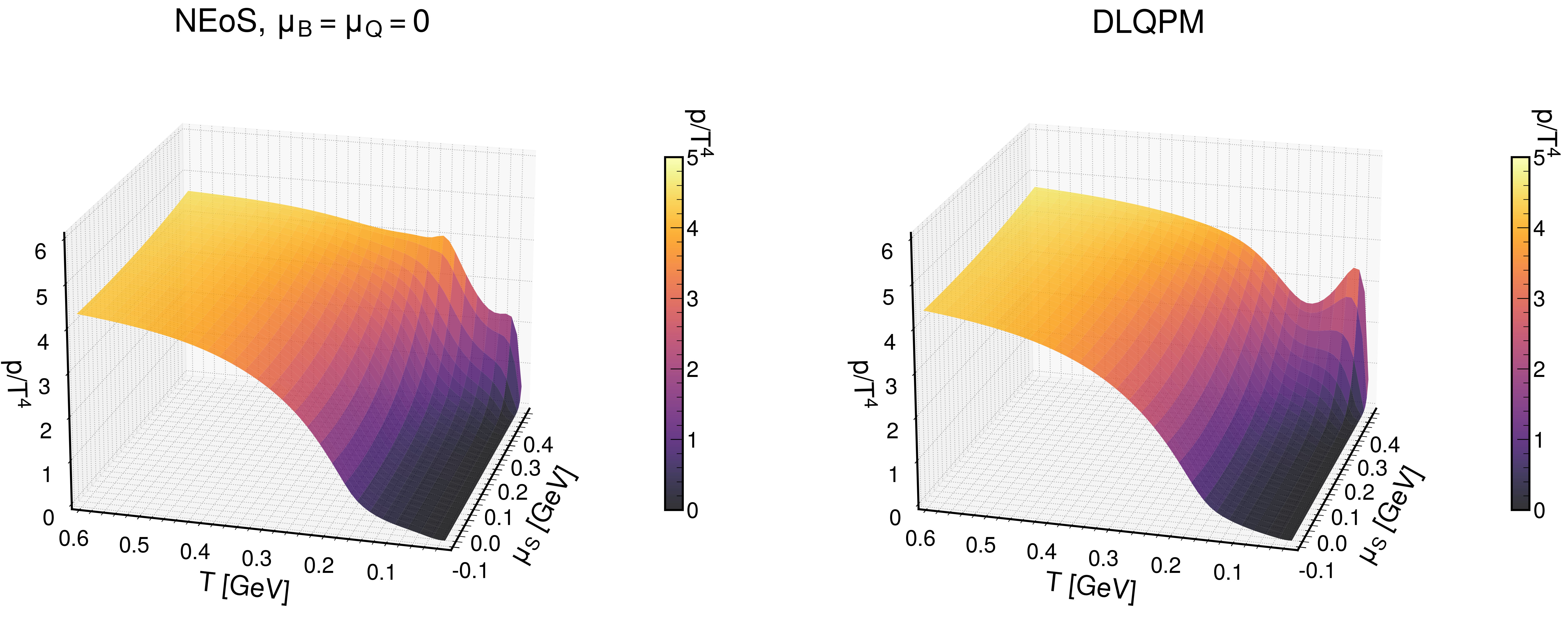}
  \caption{(Color online) The pressure $p/T^4$ across the two-dimensional planes of ($T$, $\mu_B$), ($T$, $\mu_S$), and ($T$, $\mu_Q$) compares with NEoS results~\cite{Monnai:2024pvy}.}
  \label{fig:EoS4D}
\end{figure*}

Fig.~\ref{fig:EoS4D} shows the pressure $p/T^4$ across the two-dimensional planes of ($T$, $\mu_B$), ($T$, $\mu_S$), and ($T$, $\mu_Q$), compared with NEoS results. Differences appear mainly in the region of larger chemical potential, which is expected because the DLQPM is currently constrained only by data at $\mu_B = 0$.

\end{document}